 \documentstyle[tighten,preprint,pra,aps]{revtex}
\begin{document}
\draft

\title{GRECP/MRD-CI calculations of the spin-orbit splitting
       in the ground state of Tl and of the spectroscopic properties of TlH.}
\author{A.\ V.\ Titov\cite{Email}, N.\ S.\ Mosyagin}
\address{Petersburg Nuclear Physics Institute,
         Gatchina, St.-Petersburg district 188350, RUSSIA}
\author{A.\ B.\ Alekseyev, R.\ J.\ Buenker}
\address{Theoretische Chemie, Bergische Universit{\"a}t GH Wuppertal,
         Gau{\ss}stra{\ss}e 20, D-42097 Wuppertal, GERMANY}
\date{\today}
\maketitle

\begin{abstract}
 The generalized relativistic effective core potential (GRECP) approach is
 employed in the framework of multireference single- and double-excitation
 configuration interaction (MRD-CI) method to calculate the spin-orbit (SO)
 splitting in the $^2P^o$ ground state of the Tl atom and spectroscopic
 constants for the $0^+$ ground state of TlH. The 21-electron GRECP for Tl is
 used and the outer core $5s$ and $5p$ pseudospinors are frozen with the help
 of the level shift technique. The spin-orbit selection scheme with respect to
 relativistic multireference states and the corresponding code are developed
 and applied in the calculations. In this procedure both correlation and
 spin-orbit interactions are taken into account. A [4,4,4,3,2] basis set is
 optimized for the Tl atom and employed in the TlH calculations. Very good
 agreement is found for the equilibrium distance, vibrational frequency, and
 dissociation energy of the TlH ground state ($R_e=1.870$ \AA, $\omega_e=1420$
 cm$^{-1}$, $D_e=2.049$ eV) as compared with the experimental data ($R_e=1.868$
 \AA, $\omega_e=1391$ cm$^{-1}$, $D_e=2.06$ eV).

{\bf SHORT NAME:} GRECP/MRD-CI calculations on Tl and TlH

{\bf KEYWORDS FOR INDEXING:} Relativistic Effective Core Potential
 Configuration Interaction, Molecule with heavy atoms,
 Electronic structure calculation.

\end{abstract} 

\section{Introduction}
 \label{Intro}



 During the last few years a large number of publications have dealt with
 calculations of the $^2P^o_{1/2} - {^2}P^o_{3/2}$ splitting in the ground
 state of the Tl atom and spectroscopic constants for the $0^+$ ground state of
 TlH.  Such interest to these systems arises because of their relatively simple
 electronic structure in the valence region. This makes them very convenient
 objects for testing methods for the description of relativistic and
 correlation effects. We can mention some recent papers
 \cite{RakowitzTl,DzubaTl,WahlgrenTl,LeiningerTl,BuenkerTl,EliavTl,LandauTl} in
 which the electronic structure of thallium was studied and papers
 \cite{RakowitzTlH,DiLabio,HSLee,YKHan} in which the calculation of
 spectroscopic constants for TlH was carried out.
 With the exception of the atomic RCC calculation by Eliav {\it et al.}
 \cite{EliavTl,LandauTl} and the atomic CI/MBPT2 calculation by Dzuba {\it et
 al.} \cite{DzubaTl}, the published results cannot be considered to be very
 accurate and reliable, however, primarily because of the rather small basis
 sets and the small numbers of correlated electrons.

 In calculations of Tl and TlH with the use of the relativistic effective core
 potential (RECP) approximation~\cite{ObzorECP}, in which only 13 thallium
 electrons are treated explicitly (13e-RECPs), one more problem appears. The
 correlation of the outer core (OC) and valence (V) electrons, occupying the
 $5d$ and $ns,np,nd$ ($n = 6, 7,\dots$) orbitals, respectively, cannot be
 satisfactorily described, mainly because the smoothed V-pseudoorbitals
 (pseudospinors) have the wrong behaviour in the OC region. One-electron
 functions $\phi^{corr}_{x,k}(r)$, being some linear combinations of virtual
 orbitals, correlate to occupied orbitals $\phi^{occ}_x$ (where $x=c,v$ stands
 for the OC and V orbital indices) and are usually localized in the same space
 region as $\phi^{occ}_x$. Therefore, the original ``direct'' Coulomb
 two-electron integrals describing the OC-V correlation of $\phi^{occ}_c$ and
 $\phi^{occ}_v$ can be well reproduced by those with the pseudoorbitals,
 despite their localization in different space regions.  However, a
 two-electron integral describing the ``exchange'' part of the OC-V
 correlation,

\begin{equation}
   \int_{\bf r}  d{\bf r}\
        \phi_{c,k'}^{{corr}^{\dagger}}({\bf r})\phi^{occ}_v({\bf r})\
   \int_{\bf r'} d{\bf r'}\
        \phi_{v,k}^{{corr}^{\dagger}}({\bf r'})\phi^{occ}_{c}({\bf r'})\
   \frac{1}{{\bf r} - {\bf r'}}\ ,
 \label{2elInt}
\end{equation}
 cannot be well reproduced because the V-pseudoorbitals are smoothed in the OC
 region where the OC-pseudoorbitals are localized (for more theoretical
 details, see Ref.~\cite{GRThGr}).

 The first RECPs for Tl with the $5s, 5p$ shells treated explicitly
 (21e-RECPs) for which this disadvantage of the earlier ``semicore'' RECPs was
 overcome were generated and tested in single-configurational calculations by
 Mosyagin {\it et al.}~\cite{Mosyagin2,Tupitsyn1}. Some other inherent problems
 of the ``nodeless'' RECPs were also solved with the 21-electron Generalized
 RECP (21e-GRECP) version presented in Ref.~\cite{Mosyagin2,Tupitsyn1}.  In
 Ref.~\cite{Tupitsyn1}, for the case of the 21e-GRECP it was also shown that
 the $5s, 5p$ pseudospinors could be frozen while still providing significantly
 higher accuracy than 13e-RECPs because the valence and virtual $ns$ and $np$
 ($n = 6,7,\dots$) pseudoorbitals in the former case already have the proper
 nodal structure in the OC region.

\section{The GRECP operator in the spin-orbit representation}
 \label{Sec-GRECP}

 In most existing quantum-chemical codes for molecular calculations with RECPs
 (as well as in the MRD-CI code used in the present work) spin-orbit basis sets
 are used. In these versions the number of the two-electron integrals is
 substantially smaller than in the case of spinor basis sets providing the same
 level of correlation treatment. Spin-orbit basis sets are preferable in the
 calculations in which correlation effects give a higher contribution to the
 properties of interest than those of a relativistic nature. This is usually
 the case for valence and outermost core electrons, which mainly determine
 chemical and spectroscopic properties of molecules.

 Together with the spin-orbit basis set, the GRECP for Tl should also be
 employed in the spin-orbit representation. Following
 Ref.~\cite{Ermler,Hafner}, the components of the spin-averaged part of the
 GRECP operator called the averaged relativistic effective potentials (AREP)
 are written in the form \cite{GRThGr,Mosyagin2}:

\begin{equation}
   U_{n_vl}^{AREP}(r) = \frac{l+1}{2l+1} U_{n_vl+}(r)
		    + \frac{l}{2l+1} U_{n_vl-}(r),
 \label{Un_vl}
\end{equation}

\begin{equation}
   {\bf U}_{n_cl}^{AREP}(r) = \frac{l+1}{2l+1}
                      {\bf V}_{n_cn_vl+}(r)
                    + \frac{l}{2l+1} {\bf V}_{n_cn_vl-}(r),
 \label{Un_cl}
\end{equation}

\begin{eqnarray}
   {\bf V}_{n_cn_vl\pm}(r) & = & \bigl[ U_{n_cl\pm}(r)-U_{n_vl\pm}(r) \bigr]
                               \widetilde{\bf P}_{n_cl\pm}(r)
                               +\widetilde{\bf P}_{n_cl\pm}(r)
                               \bigl[U_{n_cl\pm}(r)-U_{n_vl\pm}(r)\bigr]
                         \nonumber\\
                         & - & \sum_{n_c'} \widetilde{\bf P}_{n_cl\pm}(r)
                               \biggl[\frac{U_{n_cl\pm}(r)+U_{n_c'l\pm}(r)}{2}
                               -U_{n_vl\pm}(r)\biggr]
                                \widetilde{\bf P}_{n_c'l\pm}(r)\ ,
 \label{Vn_cn_vlj}
\end{eqnarray}
 where $U_{nl\pm}(r)$ are the potentials generated for the
 $\tilde{\varphi}_{nl\pm}(r)$ pseudospinors by means of the Goddard
 scheme~\cite{Goddard}; $n_x$ is the principal quantum number of an outercore
 ($n_c$), valence ($n_v$) or virtual ($n_a$) pseudospinor; $l$ and $j$ are
 angular and total electron momenta; $\pm$ stands for $j=l\pm 1/2$;
 $\widetilde{\bf P}_{n_cl\pm}(r)$ is the radial projector on the OC
 pseudospinors:

\begin{equation}
  \widetilde{\bf P}_{n_cl\pm}(r) =
     \sum_m \widetilde{|n_c,l,\pm,m\rangle} \widetilde{\langle n_c,l,\pm,m|}\ .
 \label{Pnclj}
\end{equation}
 Clearly, the AREP component of the GRECP may be used in calculations with
 nonrelativistic quantum-chemical codes in order to take account of
 spin-independent relativistic effects.

 The operator of the effective spin-orbit interaction can be derived following
 the expression for the spin-angular projector ${\bf P}_{l\pm}$
 from Ref.~\cite{Hafner}:
\begin{equation}
        {\bf  P}_{l\pm}(\Omega,\sigma)\
         =\ \frac{1}{2l+1} \Bigl[ \Bigl(l+\frac{1}{2}\pm \frac{1}{2}\Bigr)
           {\bf  P}_l(\Omega) \pm
          2 {\bf P}_l(\Omega)
          \vec{\bf l}\vec{\bf s} {\bf P}_l(\Omega) \Bigr]\ .
 \label{Oper_Pnljs}
\end{equation}
 Its components, called the effective spin-orbit potentials (ESOP), can be
 written as~\cite{GRThGr,Mosyagin2}

\begin{equation}
   \Delta U_{n_vl}(r) = U_{n_vl+}(r) - U_{n_vl-}(r),
\end{equation}

\begin{equation}
   \Delta {\bf U}_{n_cl}(r) =
                      {\bf V}_{n_cn_vl+}(r) -
                      {\bf V}_{n_cn_vl-}(r),
\end{equation}

\begin{equation}
   {\bf U}_{nl}^{ESOP} = \frac{2\Delta {\bf U}_{nl}(r)}{2l+1}
   {\bf P}_{l} \vec{\bf l}\vec{\bf s}\ ,
\end{equation}

\begin{equation}
   {\bf P}_{l} = \sum_{m=-l}^{l} | lm \rangle \langle  lm |\ ,
\end{equation}
 where $|lm\rangle \langle lm|$ is the projector on the spherical function
 $Y_{lm}$.

 Neglecting the difference between $U_{n_vL}^{AREP}$ and $U_{n_vLJ}$
 for virtual pseudospinors with $l>L$ (for theoretical details see
 Ref.~\cite{GRThGr}), one can write the GRECP operator ${\bf U}$ as
\begin{eqnarray}
 {\bf U} & = & U_{n_vL}^{AREP}(r) + \sum_{l=0}^{L-1}
                   \Bigl[U_{n_vl}^{AREP}(r)-U_{n_vL}^{AREP}(r)\Bigr]
                   {\bf P}_{l}   \nonumber\\
             & + & \sum_{n_c}\sum_{l=0}^{L} {\bf U}_{n_cl}^{AREP}(r)
                   {\bf P}_{l} +
                    \sum_{l=1}^{L} \Bigl[{\bf U}_{n_vl}^{ESOP} +
                    \sum_{n_c}{\bf U}_{n_cl}^{ESOP}\Bigr]
                    {\bf P}_{l}\ .
 \label{GRECP_LS}
\end{eqnarray}

 Note that the nonlocal terms with the projectors on the most important
 correlation functions
 $\tilde{\varphi}_{n_x l\pm; n_k l_k \pm}^{corr}(r)$ (\ref{2elInt}) (where
 $x=c,v$) localized mainly in the OC and V regions and with the corresponding
 potentials $U^{corr}_{n_kl_k\pm}(r)$
 can be taken into account in the considered expressions for the GRECP
 operator additionally to those with the OC projectors.
 Obviously, the non-local GRECP terms for the frozen OC pseudospinors can be
 omitted in the sum over ($n_cl$) in Eq.~(\ref{GRECP_LS}).

 We should emphasize that in spite of the rather complicated form of the above
 GRECP operator, the main computational effort in calculating matrix elements
 with the GRECP is caused by the standard radially-local operator, which is
 also a part of conventional RECP operators, and not by the non-local GRECP
 terms. Thus, the additional complications in calculations with GRECPs are
 negligible in comparison with treatments employing conventional semi-local
 RECPs if comparable gaussian expansions are used for the partial potentials.
 The more critical point is that the effort in the calculation and
 transformation of two-electron integrals is {\it always} substantially higher
 than that in the computation of RECP integrals for all known RECP versions
 (including GRECPs) when appropriately large basis sets are employed in the
 precise calculations.

\section{Frozen-core approximation for the outer-core shells}
 \label{Sec-OC_Fr}


 To perform precise calculations of chemical and spectroscopic properties,
 correlations should be taken into account not only within the valence regions
 of heavy atoms and heavy-atom molecules but in the core regions and between
 the valence and core electrons as well. In practice, the goal is to achieve a
 given level of accuracy by correlating as small a number of electrons as
 possible, thus reducing the computational effort. However, as discussed in the
 Introduction, the accuracy of the RECPs generated for a given number of
 explicitly treated electrons cannot always satisfy the accuracy requirements
 expected from correlating all these electrons in the corresponding
 all-electron calculation. This is true, in particular, for calculations of Tl,
 having a $5d^{10} 6s^2 6p^1$ leading configuration in the ground state, and
 its compounds.

 To attain an accuracy level of 400 cm$^{-1}$ for the $^2P^o_{1/2} -
 {^2}P^o_{3/2}$ splitting in the ground state and for excitation energies to
 low-lying states of Tl and to take account of the core polarization, one
 should correlate at least 13 electrons, i.e. include the $5d$ shell. This is
 achieved in the present MRD--CI calculations with $f$ and $g$ basis functions
 describing mainly polarization of the $5d$ shell (for other recent results
 see, e.g., \cite{RakowitzTl,WahlgrenTl,LeiningerTl}). Some data from our
 13e-CI calculations of the SO-splitting in the ground state of Tl are
 collected in Table~\ref{Tl-CI} in comparison with the 3e-CI results, which in
 our DF/CI (Dirac-Fock calculations followed by CI) and the GRECP/CI
 calculations have errors of about 600 cm$^{-1}$.

 We also should mention the recent relativistic coupled-cluster (RCC) results
 of Landau {\it et al.}~\cite{LandauTl}, in which 35 electrons are correlated
 and a decrease of close to 90 cm$^{-1}$ in the above mentioned SO splitting is
 due to the Breit interaction. Note that this interaction is not yet taken into
 account in the RECPs considered in the present work.

 Obviously, the $5d$ shell should also be explicitly treated in calculations
 of molecules containing Tl to take into account core relaxation and
 polarization effects with satisfactory accuracy. For these calculations it
 would be optimal to use the RECPs with 13 electrons of Tl treated explicitly
 (13e-RECPs) such as the RECP of Ross {\it et al.}~\cite{RossCsRn} or our
 valence RECP version \cite{Tupitsyn1}. None of the known nodeless 13e-RECPs
 can provide the aforementioned accuracy, however. Although
 single-configurational tests \cite{Mosyagin2,Tupitsyn1} give errors of 100
 cm$^{-1}$ or somewhat more for excitation energies to low-lying states, they
 are dramatically increased for 13e-RECPs if all 13 electrons are correlated.
 The reasons are discussed in the Introduction (one can also see the results of
 the 13e-RECP/MRD-CI calculations in Ref.~\cite{BuenkerTl} and of the
 13e-PP/MRCI calculations in Ref.~\cite{LeiningerTl}).

 To overcome this disadvantage, one should use RECPs with at least 21
 electrons, e.g. 21e-GRECP~\cite{Mosyagin2,Tupitsyn1} and
 21e-PP~\cite{LeiningerTl} for Tl.  The $5s$ and $5p$ pseudospinors can be
 treated as frozen, however, while still providing the aforementioned accuracy.
 The $5p$ orbitals have energies about four times higher and their average
 radii are 1.4 times shorter than those for the $5d$ orbitals. Moreover, their
 angular correlation is supressed as compared with the $5d$ shell because the
 most important polarization functions ($5d$ for the $5p$ orbitals and $5p$
 for the $5s$ orbitals) are completely occupied in the lowest-lying states.
 Therefore, the $5s, 5p$ orbitals are substantially less active in chemical
 processes.

 In order to freeze the $5s$ and $5p$ pseudospinors, one can apply the energy
 level shift technique~\cite{GRThGr}. Following Huzinaga {\it et
 al.}~\cite{Huzinaga}, one should add the matrix elements of the SCF field
 operators (the Coulomb and spin-dependent exchange terms) over these OC
 pseudospinors to the one-electron part of the Hamiltonian together with the
 level shift terms

\begin{equation}
  \sum_{n_{c_f},l,\pm}\  B_{n_{c_f}l\pm}\ \widetilde{\bf P}_{n_{c_f}l\pm}(r)\ ,
 \label{OC_Fr-1}
\end{equation}
 where $B_{n_{c_f}l\pm}$ is at least of order $|2\varepsilon_{n_{c_f}l\pm}|$
 and $\varepsilon_{n_{c_f}l\pm}$ is the orbital energy of the OC
 pseudospinor $\widetilde{\phi}_{n_{c_f}l\pm}(r)$ to be frozen. Such nonlocal
 terms are needed in order prevent collapse of the molecular orbitals to the
 frozen states (the $5s_{1/2}, 5p_{1/2,3/2}$ pseudospinors for Tl). All terms
 with the frozen core pseudospinors described here (the Coulomb and exchange
 interactions, and the level shift operator) can easily be presented in
 spin-orbit form with the help of eq.~(\ref{Oper_Pnljs}), as was done above for
 the GRECP operator.

 More importantly, these OC pseudo{\it spinors} can be frozen in calculations
 with {\it spin-orbit} basis sets and they can already be frozen at the stage
 of calculation of the one-electron matrix elements of the Hamiltonian, as
 implemented in the MOLGEP code~\cite{MOLGEP}. Thus, any integrals with indices
 of the frozen spinors are completely excluded after the integral calculation
 step.

 In single-configurational calculations with the numerical HFJ
 code~\cite{Tupitsyn1} we have seen that the SO splitting of the $5p$ shell
 increases the resulting SO splitting of the $^2P^o$ ground state by about 400
 cm$^{-1}$, whereas the SO splitting of the $5d$ shell decreases the final SO
 splitting by almost the same value. Therefore, it is important to freeze the
 $5p_{1/2}$ and $5p_{3/2}$ (pseudo)spinors and not some averaged $5p$
 (pseudo)orbitals if the SO interaction is to be taken into account in the $5d$
 and valence shells.

 In Ref.~\cite{LeiningerTl}, the 21e-``energy-adjusted'' Pseudopotential (PP)
 having the features which have been emphasized
 \cite{GRThGr,Mosyagin2,Tupitsyn1,Titov1} as inherent for GRECPs (different
 potentials for the $5p$ and $6p$ pseudospinors in the case of Tl) is
 generated and applied to the calculation of the SO splitting in Tl, with the
 core correlations described by the core polarization potential (CPP).
 Some average OC pseudoorbitals are frozen and the SO splitting of 7810
 cm$^{-1}$ obtained in their 21e-PP/MRCI calculation is quite different than
 our result.

 After applying the projection operator of eq.~(\ref{Oper_Pnljs}) to the level
 shift\ (\ref{OC_Fr-1}), Coulomb and exchange terms with the frozen core
 pseudospinors, the AREP and ESOP parts of the GRECP operator are to be
 modified to include these new contributions. This technique was successfully
 employed in our earlier calculations of the spin-rotational Hamiltonian
 parameters in the BaF and YbF molecules~\cite{Kozlov1}.

 The freezing technique discussed above can be efficiently applied to those OC
 shells for which the spin-orbit interaction is clearly more important than the
 correlation and relaxation effects. If the latter effects are neglected
 entirely or taken into account within ``correlated'' GRECP
 versions~\cite{GRThGr}, the corresponding OC pseudospinors can be frozen and
 the spin-orbit basis sets can be successfully used for other explicitly
 treated shells. This is true for the $5p_{1/2,3/2}$ subshells in Tl, contrary
 to the case of the $5d_{3/2,5/2}$ subshells. Freezing the OC pseudospinors
 allows one to optimize an atomic basis set only for the orbitals which are
 varied or correlated in subsequent calculations, thus avoiding the basis set
 optimization for the frozen states and reducing the number of the calculated
 and stored two-electron integrals.  Otherwise, if the $5p$ shell should be
 correlated explicitly, a spinor basis set can be more appropriate than the
 spin-orbit one.

\section{The MRD-CI method}
 \label{Sec-MRDCI}

 In the multireference single- and double-excitation CI approach
 \cite{BuenkerCI}, the $\Lambda S$-basis sets of many-electron spin-adapted
 (and space symmetry-adapted) functions (SAFs) are employed. This method makes
 use of configuration selection and perturbative energy extrapolation
 techniques \cite{BuenkerCI} and employs the Table CI algorithm\
 \cite{BuenkerTCI} for efficient handling of the various open-shell cases which
 arise in the Hamiltonian matrix elements. Some new features of the selection
 scheme used in this work are considered below. The higher excitations in the
 CI treatment has been assessed by applying the generalized multireference
 analogue\ \cite{Hirsch} of the Davidson correction\ \cite{Davidson} to the
 extrapolated $T{=}0$ energies of each root.

 After selecting the $\Lambda S$-sets of SAFs for a chosen threshold $T_i\
 (i=1,2)$, they are collected together in accord with the relativistic
 double-group symmetry requirements and a spin-orbit CI (SO-CI) calculation is
 performed with these SAFs to obtain some SO-roots ($\Psi_I^{SO,T_i}$) and
 their energies (${\cal E}_I^{SO,T_i}$) which are of interest in a considered
 double group irreducible representation (irrep). Then the linear $T{=}0$
 correction is evaluated in the basis of the calculations with the $T_1$ and
 $T_2$ thresholds. Finally, the generalized Davidson (or full CI) correction is
 applied to each root of interest.

 The stage of the molecular spectroscopic constants calculation begins with
 the fitting of the relativistic CI potential curves to polynomials
 which are employed to construct appropriate Born-Oppenheimer nuclear motion
 Schr{\"o}dinger equations solved by the Dunham method with the help of the
 DUNHAM-SPECTR code of Mitin\ \cite{Mitin}.


\subsection{Features of the spin-orbit selection procedure}
 \label{Sec-SO-sel}

 Let us define a Hamiltonian {\bf H} for a molecule as

\begin{equation}
   {\bf H} = {\bf H}^{(0)} + {\bf V}^{corr} + {\bf H}^{SO} ,
 \label{H0corrSO}
\end{equation}
 where ${\bf H}^{(0)}$ is an unperturbed spin-independent Hamiltonian, ${\bf
 V}^{corr}$ is a two-electron operator describing correlations, and ${\bf
 H}^{SO}$ is a one-electron spin-orbit operator (ESOP in our case).  Let us
 choose an orthonormal basis set of SAFs $\{\Phi^{(n)\Lambda S}_I\}$ in the
 $\Lambda S$-coupling scheme (or ``spin-orbit'' basis set). In particular,
 these SAFs can be solutions of Hartree-Fock equations with a spin-averaged
 RECP for the molecule considered. The ${\bf H}^{(0)}$ Hamiltonian is
 constructed to be diagonal in the given many-electron basis set:

\begin{equation}
    {\bf H}^{(0)} \Phi^{(n)\Lambda S}_I\ =\
                      E^{(n)\Lambda S}_I \Phi^{(n)\Lambda S}_I\ ,
 \label{H0Psi}
\end{equation}
 where $n=0,1,\dots$ (see below the description of the indices in more detail).
 Additionally define ${\bf H}^{(0)}$ so that

\begin{equation}
     <\Phi^{(n)\Lambda S}_I | {\bf H}^{(0)} | \Phi^{(n)\Lambda S}_I>\
          \equiv\
         <\Phi^{(n)\Lambda S}_I | {\bf H} | \Phi^{(n)\Lambda S}_I>\
 \label{<H0>}
\end{equation}
 in order to exclude the first-order PT contributions to total energies
 of molecular states (this corresponds to the Epstein-Nesbet PT form).

 We will ignore the two-electron spin-dependent (Breit) interactions which
 ordinarily can be neglected when studying chemical and spectroscopic
 properties.  Breit and other quantum electrodynamic (QED) effects are
 relatively large for lanthanides and actinides, but for the V and OC shells
 they can be efficiently represented by the one-electron $j$-dependent RECP
 terms.

 Let us distinguish the following types of many-electron functions which are
 considered in a double-group symmetry:
\begin{itemize}
\item $ \{ \Phi^{(0)\Lambda S}_I , E^{(0)\Lambda S}_I
        \}_{I=0}^{N^{(0)\Lambda S}} $
      are reference SAFs (``Mains'') and their energies
      \begin{equation}
         E^{(n)\Lambda S}_I =
              <\Phi^{(n)\Lambda S}_I|{\bf H}^{(0)}| \Phi^{(n)\Lambda S}_I>
       \label{EnLSI}
      \end{equation}
      at $n = 0$ for those $\Lambda S$-irreps which are of interest for the
      final spin-orbit CI (SO-CI) calculation;

\item $ \{ \Psi^{(0)\Lambda S}_I , {\cal E}^{(0)\Lambda S}_I
        \}_{I=0}^{{\cal N}^{(0)\Lambda S}} $  are some of the CI solutions
      (``$\Lambda S$-roots'') and their energies
      \begin{equation}
         {\cal E}^{(0)\Lambda S}_I =
            <\Psi^{(0)\Lambda S}_I|{\bf H}^{(0)} +
                                         {\bf V}^{corr}| \Psi^{(0)\Lambda S}_I>
       \label{calEnLSI}
      \end{equation}
      in the $\Lambda S$-irrep which diagonalize the
      $({\bf H}^{(0)} + {\bf V}^{corr})$  in the subspace of Mains only;

\item $ \{ \Psi^{(0)SO}_I, {\cal E}^{(0)SO}_I \}_{I=0}^{{\cal N}^{(0)SO}} $
      are some of the SO-CI solutions (``SO-roots'' which are of interest) and
      their energies
      \begin{equation}
          {\cal E}^{(0)SO}_I =
              <\Psi^{(0)SO}_I|{\bf H}^{(0)} +
                                 {\bf V}^{corr} + {\bf H}^{SO}| \Psi^{(0)SO}_I>
       \label{EnSOI}
      \end{equation}
      which diagonalize the complete ${\bf H}$ Hamiltonian in the subspace of
      all Mains collected from all the $\Lambda S$-irreps considered;

\item $ \{ \Phi^{(1)\Lambda S}_I , E^{(1)\Lambda S}_I
        \}_{I=0}^{N^{(1)\Lambda S}} $  are the singly-excited SAFs (SE-SAFs)
      and their energies (\ref{EnLSI}) at $n=1$, i.e.\
      \begin{equation}
        \Phi^{(1)\Lambda S}_I \in \{ {\bf P}^{\Lambda S}
             {\bf a}^+_p {\bf a}_q\ \Phi^{(0)\Lambda' S'}_J \} \setminus
             \{ \Phi^{(0)\Lambda S}_K \} \qquad \forall \quad (p,q; J,K)\ ,
       \label{Phi1SOI}
      \end{equation}
      where ${\bf P}^{\Lambda S} = |\Lambda S><\Lambda S|$ is a projector on
      the subspace of the $\Lambda S$-states, ${\bf a}^+_p\ ({\bf a}_q)$
      are the creation (annihilation) operators of one-electron states
      (spin-orbitals) $\phi_p (\phi_q)$. The SE-SAFs can be automatically
      selected because of their relatively small number;

\item $ \{ \Phi^{(2)\Lambda S}_I , E^{(2)\Lambda S}_I
        \}_{I=0}^{N^{(2)\Lambda S}} $  are the doubly-excited SAFs (DE-SAFs)
      \begin{equation}
       \Phi^{(2)\Lambda S}_I \in \{ {\bf P}^{\Lambda S}
        {\bf a}^+_p {\bf a}^+_q {\bf a}_r\ {\bf a}_s\ \Phi^{(0)\Lambda' S'}_J\}
         \setminus (\{ \Phi^{(1)\Lambda S}_K \}\cup\{ \Phi^{(0)\Lambda S}_L \})
         \quad \forall \ (p,q,r,s; J,K,L)
       \label{Phi2SOI}
      \end{equation}
      and their energies (\ref{EnLSI}) at $n=2$; a SAF $\Phi^{(2)\Lambda S}_I$
      should be selected in accordance with some selection criteria to be used
      in the final SO-CI calculation. In principle, triple and higher excited
      sets of SAFs can be similarly defined.

\end{itemize}

 The correlation operator, ${\bf V}^{corr}$, has the symmetry of the molecule
 and, therefore, can be rewritten as

\begin{equation}
     {\bf V}^{corr}\ \equiv\ \sum_{\Lambda S}
                    {\bf P}^{\Lambda S} {\bf V}^{corr} {\bf P}^{\Lambda S}\ .
 \label{PVcorrP}
\end{equation}
 It normally gives the most important contribution through the second-order
 Brillouin-Wigner PT energy correction in the basis set of
 $\Phi^{(n)\Lambda S}_J$ (after appropriate redefinition of ${\bf H}^{(0)}$ in
 the subspace of Mains, see Ref.~\cite{Gershgorn,Shavitt}):

\begin{equation}
  \sum_{n=1,2} \sum_J
     \frac{|<\Phi^{(n)\Lambda S}_J|{\bf V}^{corr}|\Psi^{(0)\Lambda S}_0>|^2}
          {{\cal E}^{\Lambda S}_0 - E^{(n)\Lambda S}_J}\quad
 \label{PT2}
\end{equation}
 for the non-degenerate ground state $\Psi^{\Lambda S}_0$ with the exact
 energy ${\cal E}^{\Lambda S}_0$ in the $\Lambda S$-irrep (obviously, terms
 with $n \ge 3$ are automatically equal to zero because ${\bf V}^{corr}$ is a
 two-electron operator). A similar expression with the replacements
 $\Psi^{(0)\Lambda S}_0\to\Psi^{(0)\Lambda S}_I$ and
 ${\cal E}^{\Lambda S}_0\to{\cal E}^{\Lambda S}_I$ can be applied for excited
 states $\Psi^{\Lambda S}_I$ (some precautions should be taken concerning the
 degenerate states and the orthogonality constraints with respect to the
 lower-lying states with $J<I$).  As a result, the first rows, columns and
 energies on the diagonal of the Hamiltonian matrix

\begin{equation}
   <\Phi^{(n)\Lambda S}_J|{\bf V}^{corr}|\Phi^{(0)\Lambda S}_I> \quad , \quad
   <\Phi^{(n)\Lambda S}_J|{\bf H}|\Phi^{(n)\Lambda S}_J> \label{<Vcorr>}
\end{equation}
 for $n=1,2$ are usually employed in the selection procedures for SAFs
 $\{\Phi^{(1,2)\Lambda S}_I\}$ based on the nonrelativistic $A_k$ and $B_k$
 approximations (when ${\bf H}^{SO}$ is not taken into account)
 \cite{Gershgorn,Shavitt} or on the multi-diagonalization
 scheme~\cite{BuenkerCI} for subsequent calculations of $\Psi^{\Lambda S}_I$.
 In spite of some differences between these selection schemes, they are not
 very essential for the final CI results if a high quality reference set (set
 of Mains) and a suitably small threshold are chosen.

 For molecules with heavy and very heavy atoms, the ${\bf H}^{SO}$ operator can
 give large contributions to the energy both in second and in higher PT orders
 if a non-optimal set of Mains, $\{\Phi^{(0)\Lambda S}_I\}$, is chosen after an
 SCF calculation with the SO-averaged potentials (AREPs). The latter is the
 usual practice and the set of Mains generated in such a manner can be smaller
 than optimal for the case of large SO interaction. Therefore, not only second
 but third and maybe even higher PT order(s) can be important in the selection
 procedure for a ``bad'' set of the starting roots $\Psi^{(0)SO}_I$. This
 means that the off-diagonal matrix elements of ${\bf H}$ between secondary
 many-electron basis functions (SE-, DE-SAFs) may be introduced into the
 selection procedure because ${\bf H}^{SO}$ is a substantially off-diagonal
 operator contrary to ${\bf V}^{corr}$:

\begin{equation}
      <\Phi^{(n)\Lambda S}_I | {\bf H}^{SO} | \Phi^{(n')\Lambda' S'}_J>\
      \qquad (\Lambda S) \mbox{ and }(\Lambda' S') \mbox{ can be different,}
      \quad n' \in \{ n, |n \pm 1| \}\ .
 \label{<Hso>}
\end{equation}
 In particular, ${\bf H}^{SO}$ gives zero matrix elements between SAFs
 belonging to the same $\Lambda S$-irrep in the $D_{2h}$ or $C_{2v}$ symmetry
 groups.


 For simplicity, let us consider the selection scheme based on the $A_k$
 approximation (\ref{PT2}).

 In the nonrelativistic-type selection scheme, a SAF
 $\Phi^{(1,2)\Lambda S}_J$  is selected in a $\Lambda S$-irrep if
\begin{equation}
     \frac{|<\Phi^{(1,2)\Lambda S}_J|{\bf V}^{corr}|\Psi^{(0)\Lambda S}_I>|^2}
          {E^{(1,2)\Lambda S}_J - {\cal E}^{(0)\Lambda S}_I}\quad
          \ge\quad \delta E^{\Lambda S}_T\ ,
 \label{n<Vcorr>^2}
\end{equation}
 where $I \le N^{SO}$ and $\delta E^{\Lambda S}_T$ is a threshold criterion
 for the energy selection scheme in the ${\Lambda S}$-irrep.
 In~(\ref{n<Vcorr>^2}) we have replaced the exact ${\cal E}^{\Lambda S}_I$
 energies by the approximate ${\cal E}^{(0)\Lambda S}_I$
 values~(\ref{calEnLSI}) that corresponds to the Rayleigh-Schr\"odinger PT
 case. Such a simplification is justified for small $\delta E^{\Lambda S}_T$
 and good reference states.

 In a SO-CI calculation within some relativistic double-group irrep,
 substitutions for the reference state
 $(\Psi^{(0)\Lambda S}_I\ \to\ \Psi^{(0)SO}_I)$ and the perturbation
 $({\bf V}^{corr} \to {\bf V}^{corr} + {\bf H}^{SO})$  should be used in the
 previous expression, so that

\begin{equation}
     \frac{|<\Phi^{(1,2)\Lambda S}_J|{\bf V}^{corr} +
                                     {\bf H}^{SO}|\Psi^{(0)SO}_I>|^2}
          {E^{(1,2)\Lambda S}_J - {\cal E}^{(0)SO}_I}\quad
          \ge\quad \delta E^{SO}_T\ ,
 \label{<H-H0>^2}
\end{equation}
 where $\delta E^{SO}_T$ is a selection threshold for $\Phi^{(1,2)\Lambda S}_J$
 to be used in the subsequent SO-CI calculation.

 In more detail, the matrix element in the PT numerator of the above formula
 can be rewritten as

\begin{eqnarray}
   & |<\Phi^{(1,2)\Lambda S}_J|{\bf V}^{corr}|\Psi^{(0)SO}_I>|^2 \\
 \label{<Vcorr>^2}
   & +\ |<\Phi^{(1)\Lambda S}_J|{\bf H}^{SO}|\Psi^{(0)SO}_I>|^2 \\
 \label{<Hso>^2}
   & +\ 2\Re (<\Psi^{(0)SO}_I|{\bf V}^{corr}|\Phi^{(1)\Lambda S}_J>
           <\Phi^{(1)\Lambda S}_J|{\bf H}^{SO}|\Psi^{(0)SO}_I>)
 \label{<VcorrHso>}
\end{eqnarray}
 by taking into account eq.~(\ref{PVcorrP}) in the calculation of the matrix
 elements for ${\bf V}^{corr}$, contrary to those for ${\bf H}^{SO}$. In spite
 of mixing different ${\Lambda S}$-states due to ${\bf H}^{SO}$, the number of
 non-zero matrix elements with ${\bf H}^{SO}$ in eq.~(\ref{<VcorrHso>}) is
 usually relatively small because the SO interaction is a one-electron operator
 (see eq.~(\ref{<Hso>})) which is very localized compared with the long-range
 Coulomb interaction.  Thus, one can see that the nonrelativistic-type
 selection due to ${\bf V}^{corr}$ with respect to $\{{\bf P}^{\Lambda S}
 \Psi^{(0)SO}_I\}$ in each considered ${\Lambda S}$-irrep and automatic
 selection of all SE-SAFs $\{\Phi^{(1)\Lambda S}_J\}$ (\ref{Phi1SOI}) can be
 efficiently applied instead of eq.~(\ref{<H-H0>^2}). It must be emphasized
 that contrary to the selection schemes in the nonrelativistic case, SE-SAFs
 should be generated with respect to the Mains from all the used ${\Lambda'
 S'}$-irreps. In a more simplified treatment, the automatic selection of
 SE-SAFs can be done with respect to a subset of the most important Mains,
 e.g.\ having largest CI-coefficients in the $\Psi^{(0)SO}_I$ roots.


 Next let us consider the terms from the third-order PT energy (PT-3) for SAFs
 $\{\Phi^{(1,2)\Lambda S}_J\}$ which can be essential for the SO selection
 procedure. Below we shall discuss only matrix elements in the PT numerators of
 the corresponding PT-3 terms because specific expressions for the energy
 denominators are not essential for our analysis and conclusions.
 For simplicity, we shall omit the terms conjugate to those considered.

 The first two types of the PT-3 matrix elements are:
\begin{eqnarray}
   & <\Psi^{(0)SO}_I|{\bf H}^{SO}|\Phi^{(1)\Lambda S}_J>
          <\Phi^{(1)\Lambda S}_J|{\bf H}^{SO}|\Phi^{(1)\Lambda' S'}_K>
                   <\Phi^{(1)\Lambda' S'}_K|{\bf H}^{SO}|\Psi^{(0)SO}_I>\ ,\\
 \label{<Hso>^3}
   & <\Psi^{(0)SO}_I|{\bf H}^{SO}|\Phi^{(1)\Lambda S}_J>
           <\Phi^{(1)\Lambda S}_J|{\bf V}^{corr}|\Phi^{(1)\Lambda S}_L>
                     <\Phi^{(1)\Lambda S}_L|{\bf H}^{SO}|\Psi^{(0)SO}_I>\ .
 \label{<HsoVcorrHso>}
\end{eqnarray}
 The first intermediate state, $\Phi^{(1)\Lambda S}_J$, is a test SE-SAF and
 the indices for other intermediate SAFs run over all the allowed ones. The
 PT-3 terms summed over the indices of the second intermediate state give
 contributions (together with the conjugate terms) for the selection of the
 test SE-SAF. However, the SE-SAFs can be selected automatically and these
 terms are out of our particular interest.

 The following matrix element type
\begin{equation}
   <\Psi^{(0)SO}_I|{\bf V}^{corr}|\Phi^{(1,2)\Lambda S}_J>
          <\Phi^{(1,2)\Lambda S}_J|{\bf H}^{SO}|\Phi^{(1)\Lambda' S'}_K>
                 <\Phi^{(1)\Lambda' S'}_K|{\bf H}^{SO}|\Psi^{(0)SO}_I>\ .
 \label{<Hso^2Vcorr>}
\end{equation}
 can be used for the selection of $\Phi^{(1,2)\Lambda S}_J$ and
 $\Phi^{(1)\Lambda' S'}_K$ when summing over another set of intermediate
 states in the PT-3 expression. As one can see, this term can be used
 for the selection of both SE-SAFs and DE-SAFs. The above expression is
 quadratic in the (large) ${\bf H}^{SO}$ interaction contrary to the
 remaining terms considered below. The contribution of the terms with
 matrix elements~(\ref{<Hso^2Vcorr>}) can be essential and their use for
 the selection of DE-SAFs $\Phi^{(2)\Lambda S}_J$ can be important for a
 subsequent SO-CI calculation.

 The following matrix element types contain a second order perturbation in
 ${\bf V}^{corr}$ and, therefore, we can suggest that in general they are
 less important for our consideration than the above terms:

\begin{eqnarray}
   & <\Psi^{(0)SO}_I|{\bf V}^{corr}|\Phi^{(1,2)\Lambda S}_J>
          <\Phi^{(1,2)\Lambda S}_J|{\bf V}^{corr}|\Phi^{(1)\Lambda S}_L>
                <\Phi^{(1)\Lambda S}_L|{\bf H}^{SO}|\Psi^{(0)SO}_I>\ ,
 \label{<HsoVcorr^2>} \\
   & <\Psi^{(0)SO}_I|{\bf V}^{corr}|\Phi^{(1,2)\Lambda S}_J>
           <\Phi^{(1,2)\Lambda S}_J|{\bf H}^{SO}|\Phi^{(1,2)\Lambda' S'}_K>
                <\Phi^{(1,2)\Lambda' S'}_K|{\bf V}^{corr}|\Psi^{(0)SO}_I>\ .
 \label{<1VcorrHsoVcorr>}
\end{eqnarray}
 These terms, together with the conjugate ones, can be used for the selection
 of $\Phi^{(1,2)\Lambda S}_J$ and $\Phi^{(1)\Lambda S}_L$.
 The term

\begin{equation}
   <\Psi^{(0)SO}_I|{\bf V}^{corr}|\Phi^{(2)\Lambda S}_J>
           <\Phi^{(2)\Lambda S}_J|{\bf H}^{SO}|\Phi^{(2)\Lambda' S'}_K>
                     <\Phi^{(2)\Lambda' S'}_K|{\bf V}^{corr}|\Psi^{(0)SO}_I>\
 \label{<2VcorrHsoVcorr>}
\end{equation}
 can be analyzed separately because it contains both the intermediate states
 as DE-SAFs. In general, it is more difficult to take such terms into account
 in the selection procedure, because of the large number of tested DE-SAFs. We
 should note, however, that when a tested $\Phi^{(2)\Lambda S}_J$ DE-SAF is
 fixed, the other intermediate states, $\{\Phi^{(2)\Lambda' S'}_K\}$, are those
 DE-SAFs which are only singly excited with respect to the tested one.
 Therefore, the number of them will not be very high.

 For completeness, the matrix element type which is cubic in the ${\bf V}^{corr}$
 perturbation should be listed:

\begin{equation}
    <\Psi^{(0)SO}_I|{\bf V}^{corr}|\Phi^{(1,2)\Lambda S}_J>
          <\Phi^{(1,2)\Lambda S}_J|{\bf V}^{corr}|\Phi^{(1,2)\Lambda S}_K>
                  <\Phi^{(1,2)\Lambda S}_K|{\bf V}^{corr}|\Psi^{(0)SO}_I>\ .
 \label{<1Vcorr>^3}
\end{equation}
 This term is of nonrelativistic type and it is out of our particular interest
 because it does not contain the ${\bf H}^{SO}$ perturbation. Again, we can
 separate the term

\begin{equation}
    <\Psi^{(0)SO}_I|{\bf V}^{corr}|\Phi^{(2)\Lambda S}_J>
            <\Phi^{(2)\Lambda S}_J|{\bf V}^{corr}|\Phi^{(2)\Lambda S}_K>
                    <\Phi^{(2)\Lambda S}_K|{\bf V}^{corr}|\Psi^{(0)SO}_I>\ .
 \label{<2Vcorr>^3}
\end{equation}
 from the previous one only because the latter contains both the DE-SAF
 intermediate states.


 We should emphasize that the terms containing SE- or DE-SAFs in the
 intermediate states of the PT-3 expressions are not taken into account in the
 $B_k$ and multi-diagonalization selection procedures, although these schemes
 include, in fact, contributions of higher than the second-order PT terms.

 When analyzing the above PT-3 terms, it can be concluded that if one replaces
 the reference SO roots, $\Psi^{(0)SO}_I$, by new reference states,
 $\Psi^{(0+1)SO}_I$, which diagonalize the complete Hamiltonian ${\bf H}$ for
 the sets of both Mains and SE-SAFs taken together, and applies the selection
 criterion based on the second-order PT~(\ref{<H-H0>^2}), then the main part of
 the above PT-3 terms will be taken into account in such a selection. An
 exception occurs for terms (\ref{<2VcorrHsoVcorr>}) and (\ref{<2Vcorr>^3}),
 but in general they are thought to be less important than the other
 third-order PT terms.

 In a more sophisticated treatment, the reference $\Psi^{(0+1')SO}_I$ SO
 states can be generated when diagonalizing ${\bf H}$ for the sets of Mains and
 those SE-SAFs ($\{\Phi^{(1')\Lambda S}_J\}$), which are automatically
 generated with respect to the most important subset of Mains
 ($\{\Phi^{(0')\Lambda S}_I\}$). The latter subset can be selected from a
 preliminary CI calculation for the set of Mains, e.g. in a basis of
 configurations with the highest CI coefficients in $\Psi^{(0)SO}_I$, and so
 on. This is worthwhile in order to reduce the number of SAFs in the resulting
 reference states $\Psi^{(0+1')SO}_I$ rather than in $\Psi^{(0+1)SO}_I$, thus
 reducing the selection time which can otherwise be very large.

 We should also note that the trial SE- and DE-SAFs, which are tested in the
 above selection procedure, are generated only for the set of Mains and not for
 the $\{\Phi^{(0+1')\Lambda S}_I\}$ set. Therefore, the number of the tested
 configurations and the selection time are reasonably limited. If the number
 of configurations used in $\{\Phi^{(0+1')\Lambda S}_I\}$ is not high, one can
 extend the set of Mains by including the above subset of SE-SAFs, thus
 obviously enlarging the set of the consequently generated and tested SE- and
 DE-SAFs.

 Again we should emphasize that it is not necessary to use the third-order PT
 or the suggested automatic selection of SE-SAFs in a selection procedure if a
 fairly good set of the reference roots $\Psi^{(0)SO}_I$ is used, i.e., if they
 provide good approximations to the required solutions $\Psi^{SO}_I$. In
 particular, if the $\{\Psi^{(0)SO}_I\}$ set is obtained from a preliminary
 series of SO-CI calculations of the studied states, this can be superfluous.


 As an alternative to the above selection schemes with respect to the PT
 energy, the PT expressions for the CI coefficient of a trial SE- or DE-SAF can
 be also explored. Applying the above PT analysis to the case of the
 $\Psi^{(0+1')SO}_I>$ reference state, a $\Phi^{(1'',2)\Lambda S}_J$ SAF is
 selected if its CI coefficient $C^{(1'',2)\Lambda S}_J$ satisfies the
 inequality

\[
   |C^{(1'',2)\Lambda S}_J| \ge C_{min}\ ,
\]
 where $C_{min}$ is the selection threshold for the CI coefficients and

\begin{equation}
  C^{(1'',2)\Lambda S}_J =
     \frac{<\Phi^{(1'',2)\Lambda S}_J|{\bf H}|\Psi^{(0+1')SO}_I>}
          {{\cal E}^{(0+1')SO}_I - E^{(1'',2)\Lambda S}_J}\ ,
 \label{<C_J>}
\end{equation}
 is the first-order PT value for the CI coefficient of a tested SAF which is
 not included in the subset of the $\Phi^{(1')\Lambda S}_J$ reference SE-SAFs.

 Such a means of selection can be preferable if those properties of primary
 interest cannot be calculated from potential energy curves or surfaces.
 Moreover, the PT selection with respect to both the energy and the CI
 coefficients can be applied simultaneously if the properties are of different
 nature.

\section{Calculations}
 \label{Sec-Calc}

 In the CI calculations of Tl and TlH we used the MRD-CI package\
 \cite{BuenkerCI} combined with the SO selection codes based on the scheme
 described above. Our test calculations have shown that spin-orbit selection is
 very helpful for preparation of appropriate sets of Mains and for reducing
 effort in the final CI calculations with an optimal set of selected SAFs.

\subsection{Spin-orbit splitting in the ground state of Tl}
 \label{Sec-Tl}

 Calculations for the Tl atom were performed to optimize the basis set and the
 level shift GRECP parameters for the 21e/8fs-GRECP, i.e.\ the 21 electron
 GRECP with 8 electrons occupying the frozen OC pseudospinors, $5s_{1/2}$ and
 $5p_{1/2,3/2}$. The quality of the generated basis set is analyzed by
 calculating the ${^2}P^o_{1/2} - {^2}P^o_{3/2}$ splitting for the ground
 state.

 Before discussing the present results, it is worthwhile to to make some brief
 comments concerning numerous values for the Tl(${^2}P^o$) spin-orbit splitting
 calculated and published in the last years. It is well known that this
 quantity calculated at the one-configuration Dirac-Fock level agrees very
 well, within 100 cm$^{-1}$, with the experimental value of 7793 cm$^{-1}$\
 \cite{Moore}. However, such a good agreement results from the fortuitous
 cancelation of a number of large errors caused by the DF approximation. The
 situation changes dramatically when even the three outermost $6s^26p^1$
 electrons are correlated. Some Tl(${^2}P^o$) splitting values from our 3e-CI
 calculations employing different codes and basis sets are given in
 Table~\ref{Tl-CI} and lie between 7130 and 7210 cm$^{-1}$. The corresponding
 3e-CI results obtained by other groups after 1996 range from 6800 to 7800
 cm$^{-1}$ and such a large divergency can not be considered as satisfactory,
 because the ground state of Tl has a very simple configuration structure as
 compared to other heavy elements. We consider our calculated values of about
 7200 cm$^{-1}$ for this splitting as reliable for an approach in which the
 $5d$ spinors are frozen after the DF calculation of the nonrelativistically
 averaged $6s^26p^1$ configuration. Taking into account that a contribution of
 aproximately -100 cm$^{-1}$ arises from the Breit terms, the deviation from
 the experimental value for the splitting is around 600--700 cm$^{-1}$ (this
 size of an error can be justified theoretically). As will be shown below, the
 computated value for the Tl(${^2}P^o$) splitting can be significantly improved
 if $5d$ electrons are explicitly included in the calculations and the
 corresponding basis set contains functions with sufficiently high angular
 momenta.

 The optimal basis set was selected in a series of MRD-CI calculations for Tl
 (with different sets of primitives and numbers of contracted $s,p,d,f$ and $g$
 functions) to minimize the sum of energies for the ground $^2P_{1/2}$ and
 ${^2}P_{3/2}$ states. In these calculations, the SAFs were selected in the
 $^2B_{1u}$, $^2B_{2u}$, and $^2B_{3u}$ irreps of the $D_{2h}$ group
 (nonrelativistic-type degenerate $^2P$ ground states belong to these irreps)
 because these doublets are strongly mixed by the SO interaction, resulting in
 the splitting of the ground $^2P$ state. We have found that two $g$ functions
 should be added to the basis set, giving a contribution of about 9000
 cm$^{-1}$ to the $^2P^o$ ground state total energies. The resulting
 [4,4,4,3,2] basis set and GRECP parameters for Tl can be found on {\it
 http://www.qchem.pnpi.spb.ru}.

 For the [4,4,4,3,2] basis set we have also performed MRD-CI calculations
 including SAFs from the $^2A_u$ irrep and SAFs with quartet multiplicity
 ($^4B_{1u}$, $^4B_{2u}$, $^4B_{3u}$, and $^4A_u$). In our calculations with
 different basis sets, their contributions have decreased the SO splitting by
 about 170 cm$^{-1}$ and the total energy by about 2000 cm$^{-1}$. One can see
 from Table~\ref{Tl-CI} that this decrease is mainly caused by the $p$- and
 $d$-components which arise from reexpansion of the leading spinor
 configuration in terms of the spin-orbit configurations. For good accuracy we
 can recommend the inclusion of $\Lambda'|S\pm 1|$-irreps for the calculation
 of states having leading configurations in $\Lambda S$-irreps.

 In Table~\ref{Tl-CI} some of our final MRD-CI results are collected together
 with the atomic relativistic coupled-cluster (RCC) results~\cite{EliavTl}
 obtained with a very large basis set.
 In these MRD-CI calculations altogether 627 Mains in three basic irreps and
 about 100 Mains in five additional irreps were involved and SE-SAFs were
 automatically generated for three Mains to prepare the reference
 $\{\Psi^{(0+1')SO}_I\}_{I=1}^3$ states. Relatively small thresholds,
 $T_1$=0.03 and $T_2$=0.01 $\mu E_h$, are used in the final runs with the
 [4,4,4,3,2] basis set (for the $T$=0 threshold and full-CI extrapolations),
 thus selecting respectively about 190000 and 450000 SAFs altogether.

 One can see that the best SO splitting calculated in the present work
 underestimates the experimental result~\cite{Moore} by about 400 cm$^{-1}$
 (recall that additionally about 90 cm$^{-1}$ is due to the Breit
 interaction~\cite{LandauTl}).  Analyzing our previous GRECP/RCC calculations
 of Hg~\cite{MosyaginHg} it can be concluded that this occurs due to the
 neglect of the OC-V correlations with the OC $5p$ and $4f$ shells, and to a
 lesser extent with $5s$ rather than due to the atomic basis set
 incompleteness, the GRECP errors or the restricted CI approximation. The OC-V
 correlation (contribution to the total energy) in Tl and Hg will have the same
 order of magnitude for respective pairs of correlated electrons (spinors).

 We also studied the reliability of the linear $T{\to}0$ extrapolation
 procedure currently used in the MRD-CI code. In the final results of our
 MRD-CI calculations the corresponding correction gives the highest
 contribution to the cumulative error. So this is a bottleneck of the present
 Tl and TlH calculations with the large number of Mains.

\subsection{Spectroscopic constants of the ground state in TlH}
 \label{Sec-TlH}

 The explicit treatment of $5d$ electrons in precise TlH (TlX) calculations is
 necessary not only due to the strong correlation between these and the valence
 electrons of Tl, but also because of the substantial influence of
 relaxation-polarization effects in this shell on the bond formation. This
 cannot be very accurately taken into account by employing a polarization
 potential \cite{PolarizPot,Leininger} in combination with, e.g., 3e-RECPs
 \cite{Mosyagin2,Tupitsyn1}. The influence of other atoms (X) in a TlX molecule
 on the $5p$, $5s$ and $4f$ shells of Tl is significantly smaller and can be
 neglected if an accuracy of a few hundreds of wavenumbers for the excitation
 energies of low-lying states is sufficient. We neglected their contributions
 in calculation of the TlH spectroscopic constants.

 In calculating spectroscopic properties for the TlH ground state
 (Table~\ref{TlH-13eCI}) we used the contracted [4,4,4,3,2] basis set for
 thallium discussed above and the [4,3,1] set for hydrogen (see~{\it
 http://www.qchem.pnpi.spb.ru}) contracted from the primitive (6,3,1) gaussian
 basis set of Dunning~\cite{Dunning}. The SAFs were selected in the $^1A_1$,
 $^3B_1$, $^3B_2$ and $^3A_2$ $\Lambda S$-irreps of the $C_{2v}$ group because
 the triplet states are most strongly admixed by the SO interaction to the
 nonrelativistic $^1A_1$ (or $^1\Sigma^+$ in $C_{\infty v}$) ground state
 producing the relativistic $0^+$ ground state in the double $C_{\infty v}^*$
 group.

 We have performed three series of TlH calculations for 16 interatomic
 distances. In these runs, the $\Psi^{(0+1')SO}_0$ reference SO states are
 generated with the MRD-CI code by diagonalizing ${\bf H}$ for the set of Mains
 and the SE-SAFs which are automatically selected with respect to the single
 configuration SCF ground state (calculated with the SO-averaged GRECP), giving
 a contribution of more than 90 \% to the final wave function.

 The first run is used for preparing an optimal set of Mains for the second
 series of SO-CI calculations. Only one SCF configuration which has the lowest
 energy in each $\Lambda S$-irrep is included into the subspace of $\Lambda S$
 Mains and, consequently, the SO reference state consists of these SCF
 configurations and the automatically selected SE-SAFs with respect to the SCF
 configuration from the $^1A_1$ irrep.


 Those SAFs were selected as Mains for the second run which had the highest CI
 coefficients in the first run. As a result, 37 Mains in all irreps together
 are employed in the second run. Relatively small thresholds, $T_1$=1.0 and
 $T_2$=0.1 $\mu E_h$, are used in the second run (for the $T{\to}0$
 extrapolation~\cite{BuenkerCI}), thus causing about 20000 and 85000 SAFs to be
 selected in the $\Lambda S$-irreps altogether.

 In the most computationally consuming third run (with the set of Mains
 consisting of the SAFs having the largest CI coefficients in the wave function
 from the second run), about 320 Mains are used altogether and the thresholds
 are set at $T_1$=0.1 and $T_2$=0.05 $\mu E_h$. About 70000 and 130000 SAFs,
 respectively, were used in the $\Lambda S$-irreps altogether in the final
 SO-CI calculations.

 One can see from Table~\ref{TlH-13eCI} that the basis set superposition error
 (BSSE) (see\ \cite{BSSE} and references) must be taken into account for an
 accurate computation of spectroscopic constants. The BSSE was studied in the
 Tl$^+$ ion calculations for the same interatomic distances as in TlH and
 estimated also in the Tl$^-$ calculations for three distances, i.e.\ with the
 ghost H atom. The same molecular basis set as in TlH was used for both the Tl
 and H atoms. The contribution from BSSE to the total energy is decisive for
 the $5d^{10}$ and $6s^2$ shells considered in the case of Tl$^+$, while its
 changing due to addition of the $6p$ electrons (which are bonding in TlH) can
 be considered as relatively small, because the difference in BSSE for Tl$^+$
 and Tl$^-$ is not significant in comparison with other errors.
 In the calculations of the spectroscopic properties with the counterpoise
 corrections (CPC), the calculated TlH points on the potential curve were
 corrected with the calculated BSSE for Tl$^+$, i.e.\ for the $5d,6s$ shells
 taken into account.

 One can see that after applying the T=0, FCI, and counterpoise corrections,
 the calculated properties are in very good agreement with the experimental
 data both in the second and third runs. The accuracy obtained is notably
 better than for other existing results for TlH (and not only for those
 presented in Table~\ref{TlH-13eCI}).
 We suggest, however, that the very good agreement of the calculated $D_e$
 with the experimental value can be fortuitous and the ``real'' (full CI) value
 can be notably different from the listed one because of the approximations
 made.

\section{Resume}
 \label{Resume}

 The SO splitting in the ground $^2P$ state of Tl is calculated by the MRD-CI
 method with the 21e-GRECP when $5d^{10}, 6s^2$ and $6p^1$ electrons are
 correlated and the $5s^2$ and $5p^6$ pseudospinors are frozen in the framework
 of the level shift technique. A [4,4,4,3,2] basis set is optimized for Tl and
 an underestimation of about 400 cm$^{-1}$ is found for the SO splitting as
 compared with the experimental data.

 Further improvement of the accuracy can be attained when correlations with
 the outer core $4f,5p$ and $5s$ shells of Tl and Breit effects are taken into
 account. We expect that this can be efficiently done in the framework of the
 ``correlated'' 21e/8fs-GRECP version in which 13 electrons are treated
 explicitly as in the present calculation. The inclusion of $h$-type functions
 is also desirable, as has been demonstrated for Hg in Ref.~\cite{MosyaginHg}.

 Fourteen electrons are correlated in the calculation of spectroscopic
 constants for the $0^+$ ground state of TlH and very good agreement with the
 experimental data is found.


 The developed spin-orbit selection scheme and code are demonstrated to be
 efficient when large sets of basis functions and reference configurations are
 required in high-precision electronic-structure calculations.

 \acknowledgments

 This work was supported by the DFG/RFBR grant N 96--03--00069 and the RFBR
 grant N 99--03--33249. AVT is grateful to REHE program of the European Science
 Foundation for fellowship grants (NN 14--95 and 22--95) to visit the
 laboratory of one of us (RJB), where part of the work was done. We are
 thankful to K.~Shulgina and T.~Isaev (PNPI) for writing some codes used for
 automatic generation of Mains.

 We are grateful to Dipl.-Ing. H.-P.~Liebermann for the help in combining the
 MOLGEP and MRD-CI codes. We are also grateful to Dr. G.~Hirsch (deceased) for
 his kind hospitality and invaluable help during visits to Wuppertal by AVT
 and NSM.

 The main part of the present calculations was carried out at the computer
 center of the Bergische Universit{\"a}t GH Wuppertal.
 JECS codes developed by PNPI quantum chemistry group were used for remote
 control of the calculations.


\newpage

\begin{table}
\begin{center}
\caption{Calculations of the spin-orbit splitting of the $^{2}P^o$ ground state in
         Tl (the $[\dots], 5d$ spinors are frozen from the SCF calculation of
         the nonrelativistically averaged $[\dots] 6s^2 6p^1$ configuration).}

\vspace{1cm}

\begin{tabular}{lcccc}
                                             &       &       &       &       \\
~~~Method                      &\multicolumn{4}{c}{SO splitting in $cm^{-1}$}\\
\hline
~~~Spinor basis sets:                        &[7,7,5]&[7,7,5,3]&[7,7,5,3,1]& \\
                                             &       &       &       &       \\
 81e-DF/3e-CI~\cite{GRThGr}                  & 7129  & 7182  & 7206  &       \\
                                             &       &       &       &       \\
 21e/18fs-GRECP/3e-CI~\cite{GRThGr}          & 7133  & 7187  & 7211  &       \\
\hline
~~~Spin-orbit basis sets:                    &[4,4,4]&[4,4,4,3]&[4,4,4,3,2]& \\
                                             &       &       &       &       \\
 21e/18fs-GRECP/3e-MRD-CI (Full CI)          &       &       &       &       \\
 (${^2}B_{1u}, {^2}B_{2u}, {^2}B_{3u}$
    irreps of $D_{2h}$)                      & 7305  & 7373  & 7398  &       \\
                                             &       &       &       &       \\
 ($\dots + {^2}A_u, {^4}B_{1u}, {^4}B_{2u},
                    {^4}B_{3u}, {^4}A_u$)    & 7133  & 7205  & 7230  &       \\
                                             &       &       &       &       \\
 21e/8fs-GRECP/13e-MRD-CI + T=0 + FCI        &       &       &       &       \\
 (${^2}B_{1u}, {^2}B_{2u}, {^2}B_{3u}$
    irreps of $D_{2h}$)                      & 7332  & 7222  & 7517  &       \\
                                             &       &       &       &       \\
 ($\dots + {^2}A_u, {^4}B_{1u}, {^4}B_{2u},
                    {^4}B_{3u}, {^4}A_u$)    & 7146  & 7044  & 7380  &       \\
\hline
~~~Spinor basis set:                         &       & & &[35,27,21,15,9,6,4]\\
                                             &       &       &       &       \\
 81e-DF/35e-RCC~\cite{EliavTl}               &       &       &       & 7710  \\
\hline
                                             &       &       &       &       \\
Experiment~\cite{Moore}                      &       &       &       & 7793  \\
\end{tabular}
\label{Tl-CI}
\end{center}
\end{table}

\hoffset=+0.0cm

\begin{table}
\begin{center}
\caption{GRECP/MRDCI calculations of the spectroscopic constants for the
         ground state of TlH.}

\vspace{1cm}

\begin{tabular}{lclc}
                                               & $R_e$ &$\omega_e$ &$D_e$  \\
Method                                         &($\AA$)&($cm^{-1}$)&($eV$) \\
\hline
SOCIEX: Tl [8,8,5,2] + H [4,3,1]               &       &           &       \\
(Rakowitz \& Marian, 1997 \cite{RakowitzTlH})  & 1.86~ &    1386   & 2.13~ \\
\hline
13e-RECP/SOCI: Tl [4,4,4,1] + H [4,2]          &       &           &       \\
(DiLabio \& Christiansen, 1998 \cite{DiLabio}) & 1.912 &    1341   & 1.908 \\
\hline
13e-REP/KRCCSD(T): Tl [4,5,5,1] + H [3,2]      &       &           &       \\
(Lee {\it et al.}, 1998 \cite{HSLee})          & 1.910 &    1360   & 2.02~ \\
21e-REP/KRCCSD(T): Tl [4,5,5,1] + H [3,2]      &       &           &       \\
(Han {\it et al.}, 2000 \cite{YKHan})          & 1.877 &           & 2.00~ \\
\hline
21e/8fs-GRECP/14e-MRD-CI Tl [4,4,4,3,2] + H [4,3,1] &  &           &       \\
(Present calculations)                         &       &           &       \\
 37 Mains, T=0.1                               & 1.858 &    1481   & 2.03~ \\
---------"--------- + CPC                      & 1.872 &    1446   & 1.984 \\
---------"--------- + T=0 + FCI                & 1.858 &    1453   & 2.10~ \\
---------"--------- + T=0 + FCI + CPC          & 1.872 &    1410   & 2.026 \\
                                               &       &        &          \\
 320 Mains, T=0.05                             & 1.866 &    1408   & 2.23~ \\
-----------"------------ + T=0 + FCI           & 1.858 &    1449   & 2.124 \\
-----------"------------ + T=0 + FCI + CPC     & 1.870 &    1420   & 2.049 \\
\hline
                                               &       &           &       \\
Experiment (Grundstr\"om \& Valberg, 1938
                             \cite{Grundstr})  & 1.866$^{\rm a}$
                                                       &    1390.7 & 2.06~ \\
Experiment (Urban {\it et al.}, 1989
                                \cite{Urban})  & 1.872$^{\rm b}$
                                                       &    1391.3 &       \\
\end{tabular}
\label{TlH-13eCI}
\end{center}
\vspace{.3cm}

\noindent $^{\rm a}$Huber \& Herzberg (1979) \cite{Huber} have published value
                   1.87 $\AA$ which can be obtained from the rotational
                   constant $B_e$.

\noindent $^{\rm b}$This value is calculated by us from $B_e$.

\end{table}

\end{document}